\def\ra{\rangle}
\def\la{\langle}
\def\be{\begin{equation}}
\def\ee{\end{equation}}
\def\ba{\begin{array}}
\def\ea{\end{array}}

\def\p{\prime}

\def\Nb{{I\!\! N}}
\def\Rb{{I\!\! R}}

\def\Cb{{\Bbb C}}
\def\cb{{\Bbb C}}

\documentclass[prl,showpacs,twocolumn,amsmath]{revtex4}
\usepackage{amssymb}
\def\qed{\leavevmode\unskip\penalty9999 \hbox{}\nobreak\hfill
     \quad\hbox{\leavevmode  \hbox to.77778em{%
               \hfil\vrule   \vbox to.675em%
               {\hrule width.6em\vfil\hrule}\vrule\hfil}}
     \par\vskip3pt}

\input amssym.def
\begin{document}
\title{Local Unitary Equivalence of Arbitrary Dimensional Bipartite Quantum States}
\author{Chunqin Zhou$^{1}$}
\author{Ting-Gui Zhang$^{2}$}
\author{Shao-Ming Fei$^{2,3}$}
\author{Naihuan Jing$^{4,5}$}
\author{Xianqing Li-Jost$^{2}$}
\affiliation{
$^1$Department of Mathematics, Shanghai Jiaotong University, Shanghai 200240, China\\
$^2$Max-Planck-Institute for Mathematics in the Sciences, 04103
Leipzig, Germany\\
$^3$School of Mathematical Sciences, Capital Normal University, Beijing
100048, China\\
$^4$School of Sciences, South China University of Technology, Guangzhou 510640, China\\
$^5$Department of Mathematics, North Carolina State University, Raleigh, NC 27695, USA}

\begin{abstract}
The nonlocal properties of arbitrary dimensional bipartite quantum
systems are investigated. A complete set of invariants under local
unitary transformations is presented. These invariants give rise to
both sufficient and necessary conditions for the equivalence of
quantum states under local unitary transformations: two density
matrices are locally equivalent if and only if all these invariants
have equal values.
\end{abstract}

\pacs{03.67.-a, 02.20.Hj, 03.65.-w}
\maketitle

{\it Introduction. ---} As a fundamental phenomenon in quantum
mechanics, the quantum nonlocality has been recently extensively
investigated. Nonlocally quantum correlated states, like quantum
entangled states \cite{EofCon} or states with nonzero quantum
discord \cite{Qdiscord}, play very important roles in many quantum
information processing such as quantum computation
\cite{DiVincenzo}, teleportation
\cite{teleport}, dense coding
\cite{dense}, cryptography \cite{crypto1} and
assisted optimal state discrimination \cite{aosd}.

Due to the fact that the nonlocal properties, e.g. the quantum
entanglement of two parts of a quantum system, remain invariant
under local unitary transformations, they can be characterized in
principle by the complete set of invariants under local unitary
transformations. For instance, the trace norms of realigned or
partial transposed density matrices in entanglement measure
\cite{norm} and the separability criteria \cite{entangle} are some
of these invariants. A complete set of invariants gives rise to full
classification of the quantum states under local unitary
transformations.

Many approaches to construct invariants of local unitary
transformations have been presented in recent years. The method
developed in \cite{Rains,Grassl}, in principle, allows one to
compute all the invariants of local unitary transformations, though
it is not easy to perform it operationally. In \cite{makhlin}, a
complete set of 18 polynomial invariants is presented for the
locally unitary equivalence of two qubit mixed states. Partial
results have been obtained for three qubits states \cite{linden},
some generic mixed states \cite{SFG, SFW, SFY}, tripartite pure and
mixed states \cite{SCFW}. The local unitary equivalence
problem for multipartite pure qubit states has been solved in
\cite{mqubit}. Then the problem for arbitrary dimensional multipartite pure states
are also solved recently \cite{mqudit}.
However, for mixed sates, generally we still have no operational
criteria to judge the equivalence of two arbitrary dimensional
bipartite states under local unitary transformations.

In this letter, we study the nonlocal properties of arbitrary
dimensional bipartite quantum systems and solve the local
equivalence problem by presenting a complete set of invariants such
that two density matrices are locally equivalent if and only if all
these invariants have the equal values in these density matrices.
These invariants can be explicitly calculated and give rise to an
operational way to judge the local unitary equivalence for
nondegenerate density matrices. For degenerate case, due to the eigenvector decompositions
of a given state are not unique in the degenerate eigenvectors' subspace, the approach is
no longer operational since the expressions of the set of invariants are not unique.
Nevertheless the set of invariants is still complete in the sense that if, in a suitable
eigenvector decompositions, two states have the same values of the invariants, they must
be equivalent under local unitary transformations. In particular, we also present
a set of invariants that are independent on the detailed eigenvector decompositions
of a quantum state. These invariants present a necessary criterion on local unitary
equivalence.

{\it The linear space formed by the invariants. ---} We first
establish a linear space spanned by matrices whose traces are the
invariants under local unitary transformations.

Let $H$ be an $N$-dimensional complex Hilbert space,
with $\vert i\rangle$, $i=1,...,N$, an orthonormal basis. Let $\rho$ be a
density matrix defined on $H\otimes H$ with $rank(\rho)=n\leq
N^2$. $\rho$ can be generally written as
\be\label{rho}
\rho=\sum_{i=1}^n\lambda_i|v_i\ra\la v_i|,
\ee
where $|v_i\ra$ is the
eigenvector with respect to the nonzero eigenvalue $\lambda_i$.
$|v_i\ra$ is a normalized bipartite pure state of the form:
$$
|v_i\ra=\sum_{k,l=1}^Na_{kl}^i|kl\ra,\ \ a_{kl}^i\in \Cb,\
\ \sum_{k,l=1}^Na_{kl}^ia_{kl}^{i\ast}=1,
$$
where $\ast$ stands for complex conjugation.

Two density matrices $\rho$ and $\rho^\prime$ are said to be equivalent
under local unitary transformations if there exist unitary operators
$U_1$ (resp. $U_2$) on the first (resp. second) space of $H\otimes H$ such that
\be\label{eq}
\rho^\prime=(U_1\otimes U_2)\rho(U_1\otimes U_2)^\dag,
\ee
where $\dag$ denotes transpose and complex conjugation.

To solve the local equivalence problem of the density matrices $\rho$ and $\rho^\prime$, it is sufficient
to find the complete set of invariants under local unitary transformation $U_1\otimes U_2$.

\noindent {\bf Lemma 1.} The following quantities are invariants
under local unitary transformations:
\be\label{l1} \ba{l}
J^s(\rho)=Tr_1(Tr_2\rho^s),\\[1mm]
Tr[(A_iA^{\dag}_j)(A_kA^{\dag}_l)\cdots(A_hA^{\dag}_p)],
\ea \ee
where $s=1,...,N^2$, $i,j,k,l,...,h,p=1,...,n$, $Tr_1$ and $Tr_2$
stand for the traces over the first and second Hilbert spaces
respectively,  $A_i$ denotes the matrix with entries given by $(A_i)_{kl}=a_{kl}^i$, $i=1,2,\cdots,n$.

\noindent{\sf[Proof].} Let $U_1$ and $U_2$ be unitary
transformations satisfying $U_1U_1^\dag=U_1^\dag
U_1=U_2U_2^\dag=U_2^\dag U_2=1$. Under the local unitary
transformation $U_1\otimes U_2$, we have
$\rho\to\rho^\prime=U_1\otimes U_2 ~\rho~U_1^\dag\otimes U_2^\dag$.
Correspondingly, we have $\vert\nu_i\rangle \to
\vert\nu_i^\prime\rangle=U_1\otimes U_2\vert\nu_i\rangle$, or
equivalently $A_i$ is mapped to $A_i^\prime=U_1 A_i U_2^t$, where
$U_2^t$ is the transpose of $U_2$. Therefore \be\label{aa}
A_i^\prime A_j^{\prime\dag}=U_1 A_i A_j^\dag U_1^\dag,~~~
A_i^{\prime\dag}A_j^\prime=U_2^\ast A_i^\dag A_j U_2^t \ee for any
$i,j=1,...,n$. By using relations (\ref{aa}) and taking into account
that $Tr_2|v_i\ra\la v_i|=A_iA_i^\dag$, it is straightforward to
verify that $J^s(\rho^\prime)=Tr_1[\sum_{i=1}^n\lambda_i^s
Tr_2(\vert\nu_i^\prime\rangle\langle\nu_i^\prime\vert)]
=Tr_1[\sum_{i=1}^n\lambda_i^s A_i^\prime
A_i^{\prime\dag}]=J^s(\rho)$, $Tr[(A_i^\prime
A^{\prime\dag}_j)(A_k^\prime A^{\prime\dag}_l)\cdots(A_h^\prime
A^{\prime\dag}_p)]
=Tr[U_1(A_iA^{\dag}_j)(A_kA^{\dag}_l)\cdots(A_hA^{\dag}_p)U_1^\dag]
=Tr[(A_iA^{\dag}_j)(A_kA^{\dag}_l)\cdots(A_hA^{\dag}_p)]$. Hence the
quantities in (\ref{l1}) are invariants of local unitary
transformations. \qed

If two density matrices are equivalent under local unitary
transformations, then their corresponding invariants in (\ref{l1})
must have the same values. Before proving that if two density
matrices have the same values of all the invariants in (\ref{l1}),
they must be equivalent under local unitary transformations, we
first claim that the set of independent invariants in (\ref{l1}) is
finite. In fact, each factor $A_iA_j^{\dag}$ belongs to the finite
dimensional algebra $Mat(N)$, which has a linear basis $E_{ij}$,
$i,j=1,\cdots, N$. Therefore, denoting $\tau$ the number of factors like $(A_iA^{\dag}_j)$ in
$Tr[(A_iA^{\dag}_j)(A_kA^{\dag}_l)\cdots(A_hA^{\dag}_p)]$, we have that $\tau$ is at most
$N^2$. Subjecting to the variations of
the subindices $i,j,k,l,...,h,p$, there could be many invariants of this form.
However, $\tau$ may be much less than $N^2$ for given $i,j,k,l,...,h,p$.
For instance for the case of $i=j=k=l...=h=p$, $\tau$ takes values from $2$ to $N$
(the case $\tau=1$ is trivial since $Tr(A_iA^{\dag}_i)=1$ due to the normalization
of the state $|v_i\ra$). The same is true also for the invariants
$Tr[(A^{\dag}_iA_j)(A^{\dag}_kA_l)\cdots(A^{\dag}_hA_p)]$.

Denote the subalgebra of $Mat(N)$ spanned by products of $A_iA_j^{\dag}$ by
$$
\Rb(\rho)=span\{(A_iA^{\dag}_j)(A_kA^{\dag}_l)\cdots(A_hA^{\dag}_p)\},
$$
$i,j,k,l,...,h,p=1,...,n$. Obviously  $\Rb(\rho)$ is a finite
dimensional associative algebra. Set $m=\dim \Rb(\rho)$. Let
$\{\rho_1,\rho_2,\cdots,\rho_m\}$ be a basis of the linear space
$\Rb(\rho)$.

\noindent {\bf Lemma 2.} The metric tensor matrix $\Omega$, with
entries given by $\Omega(\rho)_{ij}=Tr(\rho_i\rho_j)$,
$i,j=1,2,\cdots, m$,
is non-singular.

\noindent {\sf Proof.} We prove by contradiction. If
$\det[\Omega(\rho)]=0$, then
the $m$ row vectors of the matrix $\Omega(\rho)$ are linear dependent.
There exist $c_1,c_2,\cdots,c_m\in \cb $ which are not all zero, such that
$$
\sum_{i=1}^m c_i Tr(\rho_i\rho_j)=0,~~~j=1,...,m.
$$
It follows that
\be\label{ttt}
Tr((c_1\rho_1+c_2\rho_2+\cdots+c_m\rho_m)\rho_j)=0,~~~j=1,...,m.
\ee
Since $(A_{i_1}A_{j_1}^{\dag}\cdots A_{i_r}A_{j_r}^{\dag})^{\dag}\in \Rb(\rho)$
and $\rho_i$ are linear combinations of the terms like
$A_{i_1}A_{j_1}^{\dag}\cdots A_{i_r}A_{j_r}^{\dag}$, we have
$(c_1\rho_1+c_2\rho_2+\cdots+c_m\rho_m)^\dag \in \Rb(\rho)$ and
$$
(c_1\rho_1+c_2\rho_2+\cdots+c_m\rho_m)^\dag=h_1\rho_1+h_2\rho_2+\cdots+h_m\rho_m
$$ for some $h_i\in \cb $, $i=1,2,\cdots,m$. It follows form (\ref{ttt}) that
$$
Tr[(c_1\rho_1+c_2\rho_2+\cdots+c_m\rho_m)(c_1\rho_1+c_2\rho_2+\cdots+c_m\rho_m)^\dag]=0,
$$
which implies $c_1\rho_1+c_2\rho_2+\cdots+c_m\rho_m=0$. Hence one
concludes that $\{\rho_i, i=1,2,\cdots, m \}$ are linear dependent,
which contradicts that $\{\rho_1,\rho_2,\cdots, \rho_m\}$ is a basis
of the linear space $\Rb(\rho)$. \qed

Actually, one can also prove that if $\{\rho_i, i=1,2,\cdots, m\}$
is a sequence of matrices such that $\det[\Omega(\rho)]\neq 0$, then
$\{\rho_i, i=1,2,\cdots, m\}$ is linear independent.

The invariants $Tr[(A_iA^{\dag}_j)(A_kA^{\dag}_l)\cdots(A_hA^{\dag}_p)]$ can be
equivalently written as $Tr[(A^{\dag}_iA_j)(A^{\dag}_kA_l)\cdots(A^{\dag}_hA_p)]$. Correspondingly
one has the linear space defined by
$$
\Nb(\rho)=span\{(A^{\dag}_iA_j)(A^{\dag}_kA_l)\cdots(A^{\dag}_hA_p)\},
$$
$i,j,k,l,...,h,p=1,...,n$, with finite dimension $d=\dim \Nb(\rho)$. If
$\{\theta_1,\theta_2,\cdots,\theta_d\}$ is a basis of $\Nb(\rho)$ like $\Rb(\rho)$,
we similarly have that the matrix $\Theta$ with entries given by
$\Theta(\rho)_{ij}=Tr(\theta_i\theta_j)$, $i,j=1,2,\cdots, d$,
satisfies $\det[\Theta(\rho)]\neq 0$.

\noindent {\bf Lemma 3.} If two density matrices $\rho=\sum_{i=1}^n\lambda_i\vert \nu_i\rangle\langle\nu_i\vert$ and
$\rho^\prime= \sum_{i=1}^n\lambda_i \vert\nu_i^\prime\rangle
\langle\nu_i^\prime\vert$ have the same values for the following invariants:
\be\label{v}
Tr[(A_iA^{\dag}_j)(A_kA^{\dag}_l)\cdots(A_hA^{\dag}_p)],
\ee
$i,j,k,l,...,h,p=1,...,n$, then the corresponding linear spaces
$\Rb(\rho)$ and $\Rb(\rho^\p)$ have the same dimension. Moreover,
the bases of $\Rb(\rho)$ and $\Rb(\rho^\p)$ have a one to one correspondence:
if $(A_iA^{\dag}_j)(A_kA^{\dag}_l)\cdots(A_h A^{\dag}_p)$ is a basis element of $\Rb(\rho)$,
then $(A^\p_iA^{\p \dag}_j)(A^\p_kA^{\p\dag}_l)\cdots(A^\p_h A^{\p \dag}_p)$
is a basis element of $\Rb(\rho^\p)$.
Similar results hold between the linear spaces $\Nb(\rho)$ and $\Nb(\rho^\p)$.

\noindent{\sf[Proof].} Assume that $\{\rho_\alpha, \alpha=1,2,\cdots, m\}$ is a basis of $\Rb(\rho)$.
Each $\rho_\alpha$  has a form of $(A_iA^{\dag}_j)(A_kA^{\dag}_l)\cdots(A_h A^{\dag}_p)$.
Denote $\rho^\p_\alpha=(A^\p_iA^{\p \dag}_j)(A^\p_kA^{\p \dag}_l)\cdots(A^\p_h A^{\p \dag}_p)$,
$\alpha=1,2,\cdots,m$. Since $\det[\Omega(\rho)]=\det[\Omega(\rho^\p)]\neq 0$ from Lemma 1 and the
condition (\ref{v}), we have $\{\rho^\p_\alpha, \alpha=1,2,\cdots, m\}$ is linearly independent.
Hence $\dim \Rb(\rho^\p)\geq \dim \Rb(\rho)$. In a similar way one can prove
that $\dim \Rb(\rho)\geq \dim \Rb(\rho^\p)$. Therefore $\dim \Rb(\rho^\p)= \dim \Rb(\rho)$ and
$\{\rho^\p_\alpha, \alpha=1,2,\cdots, m\}$ is a basis of $\Rb(\rho^\p)$.

Similar results between the linear spaces $\Nb(\rho)$ and $\Nb(\rho^\p)$ can be obtained
from the expression of invariants $Tr[(A^{\dag}_iA_j)(A^{\dag}_kA_l)\cdots(A^{\dag}_hA_p)]$.
\qed


{\it Local equivalence of bipartite states. ---}
We now give the necessary and sufficient condition for the equivalence of bipartite
states under local unitary transformations.

\noindent {\bf Theorem} Two arbitrary dimensional bipartite
density matrices are equivalent under local unitary transformations
if and only if there exit eigenstate decompositions (\ref{rho}) such
that the following invariants have the same values for both density
matrices:
\be\label{theorem} \ba{l}
J^s(\rho)=Tr_2(Tr_1\rho^s),~~s=1,...,N^2;\\[1mm]
Tr[(A_iA^{\dag}_j)(A_kA^{\dag}_l)\cdots(A_hA^{\dag}_p)],
\ea
\ee
where $i,j,k,l,...,h,p=1,...,n$.

\noindent{\sf[Proof].} We have proved that the quantities in (\ref{theorem}) are
invariants under local unitary transformations. We now prove that
these invariants are complete. Suppose that the states
$\rho=\sum_{i=1}^n\lambda_i\vert \nu_i\rangle\langle\nu_i\vert$ and
$\rho^\prime= \sum_{i=1}^n\lambda_i^\prime\vert\nu_i^\prime\rangle
\langle\nu_i^\prime\vert$ have the same values to the invariants in (\ref{theorem}).
From $J^s(\rho)=J^s(\rho^\prime),~~~s=1,...,N^2$, we have that
the two density matrices $\rho$ and $\rho^\prime$ have the same set
of eigenvalues, i.e. $\lambda_i=\lambda_i^\prime$ for $i=1,2,\cdots,n$.

Next we introduce the dual basis $\rho_i^*$ in $\Rb(\rho)$ such that
$Tr(\rho_i\rho_j^*)=\delta_{ij}$. In fact, let
$\Omega(\rho)^{-1}=[\Omega^{ij}(\rho)]$, then \be
\rho_i^*=\sum_{j=1}^m\Omega^{ij}(\rho)\rho_j. \ee By Cramer's rule
$\Omega(\rho)^{-1}=|\Omega(\rho)|^{-1}Adj(\Omega)(\rho)$, so
$\Omega^{ij}(\rho)$ are given by polynomials in $\Omega_{ij}(\rho)$,
consequently $\rho_i^*$ are expressed as polynomials in the basis
element $\rho_i$.

Now we study the algebra structure of $\Rb(\rho)$. First we define
the bilinear form $\langle \, , \, \rangle$ on $\Rb(\rho)$ by
\be\label{bilin}
\langle \sigma, \tau \rangle=Tr(\sigma\tau^{\dagger}),
\qquad \sigma, \tau\in \Rb(\rho).
\ee
Then $\Rb(\rho)$ becomes a
Hilbert space and each operator $\rho_i$ is bounded under the norm
$||\rho||=\langle \rho, \rho\rangle^{1/2}$. Next we have
\be\label{str1}
\rho_i\rho_j=\sum_{k=1}^m
Tr(\rho_i\rho_j\rho^*_k)\rho_k.
\ee
The invariance of the traces
(\ref{theorem}) implies that in the algebra $\Rb(\rho')$ the element
$\rho_i^{\p *}$ is given by the same formula as $\rho_i^*$:
\be
\rho_i^{\p *}=\sum_{j=1}^m\Omega^{ij}(\rho')\rho_j'=\sum_{j=1}^m\Omega^{ij}(\rho)\rho_j'.
\ee
Moreover $\rho'_i\rho'_j$ are also given by
the same structure constants: 
\be\label{str2}
\rho_i'\rho_j'=\sum_{k=1}^m
Tr(\rho_i'\rho_j'\rho^{\p\ast}_k)\rho_k' =\sum_{k=1}^m
Tr(\rho_i\rho_j\rho^*_k)\rho_k'.
\ee
Eqs (\ref{str1}) and (\ref{str2}) 
imply that the map $\rho_i\rightarrow \rho_i'$ gives an isomorphism
from the algebra $\Rb(\rho)$ onto the algebra $\Rb(\rho')$. Therefore
$\Rb(\rho)$ and $\Rb(\rho')$ are two equivalent representations of the same
underlying associative algebra. Thus there exists a
non-singular matrix $T$ such that $\rho_i=T\rho_i'T^{-1}$ for all
$i=1, \ldots, m$. In particular, we have
$A_iA_i^{\dag}=TA_i'{A'_i}^{\dag}T^{-1}$, $i=1,\ldots, m$.
As $A_iA_i^{\dag}$ are hermitian, due to the algebraic property
(\ref{bilin}) and using Theorem 12.36 in \cite{Ru} we have
\be\label{22a}
A_iA_i^{\dag}=uA_i'{A'_i}^{\dag}u^{\dag},
\ee
where $u$ is the unitary part of $T$ in the polar decomposition.


Similarly from isomorphism of the algebras $\Nb(\rho)\simeq \Nb(\rho')$,
one has
\be\label{33a}
A_i^{\dag}A_i=wA_i^{\p \dag}A'_iw^{\dag},
\ee
for some unitary matrix $w$ and all $i$.

By using relations (\ref{22a}) and (\ref{33a}) we can show
$$
A_i=uA_i'(w^*)^t,~ i=1,...,n.
$$
In fact, let $u_i$ and $u'_i$ be the unitary matrices that diagonize
the hermitian matrices $A_iA_i^\dag$ and $A_i^\dag A_i$
respectively, $u_iA_iA_i^\dag u_i^\dag
=diag\{\eta_{i_1}^2,...,\eta_{i_N}^2\}$, $u'_i A_i^\dag A_i
{u'_i}^\dag =diag\{\eta_{i_1}^2,...,\eta_{i_N}^2\}$. From the
procedure of the singular value decomposition of matrices \cite{RJ},
we have $u_i A_i {u'_i}^\dag =diag\{\eta_{i_1},...,\eta_{i_N}\}$
with $\eta_k\geq 0$. From (\ref{22a}) and (\ref{33a}), we have
$u_iA_iA_i^\dag u_i^\dag =u_iuA'_i{A'_i}^\dag u^\dag
u_i^\dag=diag\{\eta_{i_1}^2,...,\eta_{i_N}^2\}$ and $u'_i A_i^\dag
A_i {u'_i}^\dag =u'_i w{A'_i}^\dag A'_iw^\dag {u'_i}^\dag
=diag\{\eta_{i_1}^2,...,\eta_{i_N}^2\}$. Therefore from singular
value decomposition we have $u_i uA'_i w^\dag{u'_i}^\dag
=diag\{\eta_{i_1},...,\eta_{i_N}\}$. Hence we obtain
$u_iA_i{u'_i}^\dag=u_iuA'_iw^\dag{u'_i}^\dag$, i.e.
$A_i=uA'_iw^\dag=uA'_i(w^*)^t$.

Since $A_i=uA_i'(w^*)^t$, we have
$\vert\nu_i^\prime\rangle=u^\dag\otimes (w^*)^\dag
\vert\nu_i\rangle$, $i=1,...,n$, and $\rho^\prime=u^\dag\otimes
(w^*)^\dag ~\rho~u\otimes w^*$. Therefore $\rho^\prime$ and $\rho$
are equivalent under local unitary transformations. \qed

From the proof one sees that the invariants (\ref{theorem}) are complete,
finite and can be easily calculated. For nondegenerate states $\rho$ and $\rho^\prime$,
the eigenvector $\vert \nu_i\rangle$ of $\rho$ corresponds uniquely to the eigenvector
$\vert\nu_i^\prime\rangle$ of $\rho^\prime$. Hence the correspondence between $A_i$ and $A_i^\prime$
is also unique. It is straightforward to judge the local unitary equivalence
of $\rho$ and $\rho^\prime$ by simply comparing the values of invariants from $A_i$ and $A_i^\prime$
one by one.

For degenerate states, it becomes less operational
since the eigenvector decompositions (\ref{rho}) is no longer unique.
If $|v_i\rangle$, $i=1,...,r$, are the orthogonal eigenvectors with respect to a same eigenvalue.
then $\widetilde{|v_{i}}\rangle=\sum_j U_{ij}|v_j\rangle$ is
also a set of linearly independent orthogonal eigenvectors to the same eigenvalue
for any unitary matrix $(U)_{ij}=U_{ij}$.
Therefore the expressions of the invariants (\ref{theorem}) are not unique in general.
However, by linear combinations of the invariants in (\ref{theorem}) we can get
a set of invariants which does not depend on
the detailed eigenvector decompositions,
\be\label{dgi}
\ba{l}
\sum_{i,i^\prime,j,j^\prime,...,k,k^\prime=1}^r\,Tr(A_{i}A_{i^{\prime}}^{\dag}A_{j}A_{j^{\prime}}^{\dag}\cdots A_{k}A_{k^{\prime}}^{\dag}),\\[2mm]
\sum_{i,i^\prime,j,j^\prime,...,k,k^\prime=1}^r\,Tr(A_{i}^{\dag}A_{i^{\prime}}A_{j}^{\dag}A_{j^{\prime}}\cdots A_{k}^{\dag}A_{k^{\prime}}),
\ea
\ee
where the indices $\{i^{\prime},j^{\prime},...,k^{\prime}\}$ is any given permutation of
the indices $\{i,j,...,k\}$. Different given permutation gives different invariants.
Since the invariants (\ref{dgi}) are independent on the detailed eigenvector decompositions,
they gives an operational necessary condition for the local equivalence,
with respect to the subspaces spanned by the degenerate eigenvectors.
If two density matrices with the same degenerate eigenvalue are
equivalent under local unitary transformations, they must have the same values
of all the invariants in (\ref{dgi}).
As an simple example, consider $\rho=diag\{1/2,1/2,0,0\}$, $\rho^\prime=diag\{1/2,0,1/2,0\}$.
We have $A_1=A_1^\prime=\left(\ba{cc}1&0\\0&0\ea\right)$, $A_2=(A_2^{\prime})^t=\left(\ba{cc}0&1\\0&0\ea\right)$,
From (\ref{dgi}) we have that $\rho$ and $\rho^\prime$ are not local unitary equivalent, since
$Tr(A_1A_1^\dag A_1A_1^\dag+A_1A_2^\dag A_2 A_1^\dag+A_2A_1^\dag A_1A_2^\dag+A_2A_2^\dag A_2A_2^\dag)=2$, while
$Tr(A_1^\prime A_1^{\prime\dag} A_1^\prime A_1^{\prime\dag}
+A_1^\prime A_2^{\prime\dag} A_2^\prime  A_1^{\prime\dag}+A_2^\prime A_1^{\prime\dag} A_1^\prime A_2^{\prime\dag} +A_2^\prime A_2^{\prime\dag} A_2^\prime A_2^{\prime\dag})=4$.

{\it Conclusion and remarks. ---} We have investigated the
nonlocal properties of arbitrary dimensional bipartite quantum
systems and solved the local equivalence problem by
presenting a complete set of invariants such that two density
matrices are locally equivalent if and only if all these invariants
have the equal values. Although the independent invariants may vary
with the detailed bipartite states, the number of the
invariants one needs to check is finite. Here we have
dealt with the case that the dimensions of both Hilbert spaces are
the same for simplicity. Nevertheless the case that the Hilbert
spaces have different dimensions can be similarly discussed. Our
approach of constructing local invariants and algebraic proof of the
sufficiency may also shed light on multipartite case, for which only
the multipartite pure state case has been extensively studied
\cite{mqubit,mqudit}.

\noindent{\bf Acknowledgments}\, We thank K.L. Lin for useful discussions.
C.Q. Zhou thanks the Max Planck Institute for Mathematics in the
Sciences for hospitality.
This work is supported by the NSFC 10875081 and PHR201007107. N. Jing thanks the support
from Simons foundation, NSF and MPI f\"ur Mathematik in Bonn and MPI f\"ur Mathematik in Wissenschaft,
Leipzig for hospitality.

\end{document}